\documentclass[conference]{IEEEtran}
\IEEEoverridecommandlockouts
\usepackage[table]{xcolor}
\usepackage{cite}
\usepackage{graphicx}
\usepackage{lipsum}
\usepackage{stfloats}
\usepackage{bm}
\usepackage{amsmath,amssymb}
\usepackage{amsthm}
\usepackage{pifont}
\usepackage{bbding}
\usepackage{booktabs}
\usepackage{ctable}
\usepackage{threeparttable}
\usepackage{footmisc}
\usepackage{framed} 
\usepackage{array}
\usepackage{multirow}
\usepackage{longtable}
\usepackage{rotating}
\usepackage{fancyhdr}
\usepackage{lastpage}
\usepackage{layout}
\usepackage{wrapfig}
\usepackage{xcolor}
\usepackage{tabularx}
\usepackage{tabu}
\usepackage{enumitem}
\usepackage{url}
\usepackage{dsfont}
\usepackage{arydshln}

\graphicspath{{Figures/}}
\DeclareGraphicsExtensions{.png, .eps}

\makeatletter
\newcommand{\removelatexerror}{\let\@latex@error\@gobble}
\makeatother

\def\BibTeX{{\rm B\kern-.05em{\sc i\kern-.025em b}\kern-.08em
    T\kern-.1667em\lower.7ex\hbox{E}\kern-.125emX}}

\begin{document}
	\title{Hybrid Semantic/Bit Communication Based Networking Problem Optimization}
	\author{\IEEEauthorblockN{Le Xia\IEEEauthorrefmark{1},\thanks{This paper is partially funded by UK Department for Science, Innovation \& Technology (DSIT) Towards Ubiquitous 3D Open Resilient Network (TUDOR) Project, EPSRC projects,  CHEDDAR EP/X040518/1 and CHEDDAR Uplift EP/Y037421/1, and the National Science Foundation under Grants CCF-2427316 and CNS-2418308.}
				 Yao Sun\IEEEauthorrefmark{1},
				 Dusit Niyato\IEEEauthorrefmark{2},
				 Lan Zhang\IEEEauthorrefmark{3},
				 Lei Zhang\IEEEauthorrefmark{1},
				 and Muhammad Ali Imran\IEEEauthorrefmark{1}}
	\IEEEauthorblockA{\IEEEauthorrefmark{1}James Watt School of Engineering, University of Glasgow, Glasgow, UK\\
	\IEEEauthorrefmark{2}College of Computing and Data Science, Nanyang Technological University, Singapore\\
	\IEEEauthorrefmark{3}Department of Electrical and Computer Engineering, Clemson University, USA\\
	Email: Yao.Sun@glasgow.ac.uk}}

	\maketitle
	\begin{abstract}
	This paper jointly investigates user association (UA), mode selection (MS), and bandwidth allocation (BA) problems in a novel and practical next-generation cellular network where two modes of semantic communication (SemCom) and conventional bit communication (BitCom) coexist, namely hybrid semantic/bit communication network (HSB-Net).
	Concretely, we first identify a unified performance metric of message throughput for both SemCom and BitCom links.
	Next, we comprehensively develop a knowledge matching-aware two-stage tandem packet queuing model and theoretically derive the average packet loss ratio and queuing latency.
	 Combined with several practical constraints, we then formulate a joint optimization problem for UA, MS, and BA to maximize the overall message throughput of HSB-Net.
	 Afterward, we propose an optimal resource management strategy by employing a Lagrange primal-dual method and devising a preference list-based heuristic algorithm.
	 Finally, numerical results validate the performance superiority of our proposed strategy compared with different benchmarks.
	
	\end{abstract}

	\IEEEpeerreviewmaketitle
	
	\section{Introduction}
	Recent advances in semantic communication (SemCom) have attracted widespread attention, promising to significantly alleviate the scarcity of wireless resources in next-generation cellular networks.
	By leveraging cutting-edge deep learning (DL) algorithms, SemCom is capable of providing mobile users (MUs) with a variety of high-quality, large-capacity, and multimodal services, including typical multimedia content (e.g., text, image, and video streaming) and artificial intelligence-generated content (AIGC)~\cite{zhang2022toward}.
	
	Different from the conventional bit communication (BitCom) mode that aims at the precise reception of transmitted bits, SemCom focuses more on the accurate delivery of true meanings implied in source messages.
	Specifically, a semantic encoder deployed in the transmitter first filters out redundant content of the source information and extracts the core semantics that require fewer bits for transmission.
	After necessary channel encoding and decoding, the original meanings are accurately interpreted and restored from the received bits via a semantic decoder, even with intolerable bit errors in data propagation.
	Notably, either the semantic encoding or decoding is executed based on background knowledge pertinent to the delivered messages, and the higher the knowledge-matching degree between transceivers, the lower the semantic ambiguity in recovered meanings~\cite{10261329}.
	Consequently, efficient information exchanges and considerable spectrum savings can be guaranteed in SemCom.
	
	
	
	Lately, there have been some technical works on SemCom from a wireless resource management perspective.
	For instance, Zhang~\textit{et al.}~\cite{10122232} adopted a deep reinforcement learning-based dynamic resource allocation scheme to maximize the long-term transmission efficiency in task-oriented SemCom networks.
	Besides, Xia~\textit{et al.}~\cite{10261329} developed a bit-rate-to-message-rate transformation mechanism along with a semantic-aware metric called system throughput in message to jointly optimize user association (UA) and bandwidth allocation (BA) problems in SemCom-enabled networks.
	Nevertheless, there is still a missing investigation for a more practical yet novel next-generation cellular network paradigm, namely~\textit{hybrid semantic/bit communication networks} (HSB-Nets), where the two modes of SemCom and BitCom coexist.
	Note that SemCom typically requires more data processing time but produces higher semantic performance than BitCom at each transceiver, thereby determining an appropriate mode selection (MS) scheme for each MU becomes quite tricky.
	Moreover, the varying degrees of background knowledge matching among MUs can also affect the amount of bandwidth allocated by associated base stations (BSs) under different channel conditions.
	The lack of unified performance metrics for the two modes coupled with inherent practical constraints, it should be rather challenging to jointly solve three networking problems of UA, MS, and BA with the aim of overall performance optimization in a large-scale HSB-Net.

	To this end, in this paper, we systematically investigate the optimal resource management strategy in the uplink of the HSB-Net in combination with unique SemCom characteristics.
	First, we identify the unified performance metric of SemCom with BitCom as message throughput by introducing the bit-rate-to-message-rate transformation mechanism.
	Next, the knowledge matching based steady-state average packet loss ratio and queuing delay for semantic data packets in a dedicatedly devised two-stage tandem queue are theoretically derived.
	Afterward, we mathematically formulate a joint message throughput-minimization problem subject to UA, MS, and BA-related constraints.
	A Lagrange primal-dual method alongside a preference list-based heuristic algorithm is then developed to reach the optimality of wireless resource management in the HSB-Net.
	Finally, numerical results showcase the performance superiority of the proposed strategy in terms of realized overall message throughput compared with four benchmarks.
	\addtolength{\footskip}{0.02in}
    
    \section{System Model and Problem Formulation}
	
	\subsection{HSB-Net Scenario}
	Consider an HSB-Net scenario where a total of $U$ MUs and $S$ BSs are distributed, and two modes of SemCom and BitCom coexist while each MU can only select one communication mode at a time.
	Let $x_{ij}\in \{0,1\}$ denote the binary UA indicator, where $x_{ij} = 1$ means that MU $i \in \mathcal{U}=\{1,2,\cdots,U\}$ is associated with BS $j \in \mathcal{J}=\{1,2,\cdots,J\}$, and $x_{ij} = 0$ otherwise.
	Besides, we specially define the binary MS indicator as $y_{ij}\in \{0,1\}$, where $y_{ij} = 1$ represents that the SemCom mode is selected and $y_{ij} = 0$ indicates that the BitCom mode is selected.
	Meanwhile, the amount of bandwidth that BS $j$ assigns to MU $i$ is denoted as $z_{ij}$, while the total bandwidth budget of BS $j$ is assumed to be $Z_{j}$.	
	Moreover, time is equally partitioned into consecutive time slots, each with the same duration length $T$.
	Let $\gamma_{ij}(t)$ denote the signal-to-interference-plus-noise ratio (SINR) of the link at time slot $t,\ t = 1,2,\cdots,N$, which is assumed to be an independent and identically distributed (i.i.d.) random variable for different slots but remain constant during one slot~\cite{moustakas2013sinr}, having mean $\overline{\gamma}_{ij}$.
	
	\subsection{Network Performance Metric}
	Since the conveyed message itself becomes the sole focus of precise reception in SemCom rather than traditional transmitted bits in BitCom, we proceed with the performance metric developed in~\cite{10261329} to measure the overall message throughput at all SemCom-enabled MUs via employing a bit-rate-to-message-rate (B2M) transformation function.
	To be specific, the B2M function is to output the semantic channel capacity (i.e., the achievable message rate in units of messages per unit time, \textit{msg/s}) from input traditional Shannon channel capacity (i.e., the achievable bit rate in units of bits per unit time, \textit{bit/s}) under the discrete memoryless channel.
	Given this, let $\Re_{ij}(\cdot)$ denote the B2M function of the SemCom link between MU $i$ and BS $j$, its time-averaged achievable message rate is $\overline{M}_{ij}^{S}=\tau_{i}\Re_{ij}\!\left(z_{ij}\log_{2}\left(1+\overline{\gamma}_{ij}\right)\right)$, where $\tau_{i}$ ($0<\tau_i\leqslant1$) represents the average knowledge-matching degree between MU $i$ and its communication counterpart~\cite{10261329}.
	Likewise, we assume that there is an average B2M transformation ratio for each BitCom-enabled MU,\footnote{This is justified since the source-and-channel coding in BitCom typically follows prescribed codebooks, and thus the coding length can be averaged~\cite{merkle1978secure}.} to measure the network performance with a message-related metric unified with SemCom.
	Denoting the B2M transformation coefficient in the BitCom case as $\rho_{ij}$, the time-averaged achievable message rate of the BitCom-enabled link is $\overline{M}_{ij}^{B}=\rho_{ij}z_{ij}\log_{2}\left( 1+\overline{\gamma}_{ij}\right),\ 0<\rho_{ij}<1$.
	As such, if taking into account both SemCom (i.e., $y_{ij}=1$) and BitCom (i.e., $y_{ij}=0$) cases, we calculate the message rate of each link from a long-term perspective by $\overline{M}_{ij} = y_{ij}\overline{M}_{ij}^{S}+(1-y_{ij})\overline{M}_{ij}^{B}$.
	
	\subsection{Queuing Model}
	\begin{figure}[t]
		\centering
		\includegraphics[width=0.48\textwidth]{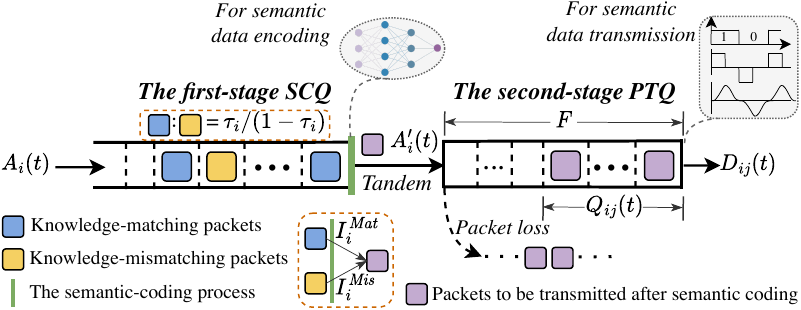} 
		\caption{The two-stage tandem queue model at each SemCom-enabled MU.}
		\label{Queuing}
    \end{figure}
	To mathematically characterize the differences between SemCom and BitCom, we exploit queuing models for uplink transmission.
	For clarity, we assume that each SemCom-enabled MU has a two-stage tandem queue, named Semantic-Coding Queue (SCQ) and Packet-Transmission Queue (PTQ), to capture the semantic packet processing queue and transmission queue, respectively.
	As for each BitCom-enabled MU, only one PTQ is considered for packet transmission.
	Note that the packets in both SCQ and PTQ are queued in a first-come-first-serve manner, while the SCQ is assumed with infinite-size memory and the PTQ has a finite buffer size $F$.
	
	A Poisson arrival process with average rate $\lambda_{i}$ (in \textit{packets/s}) of initial data packet generation is assumed for each MU $i$ ($\forall i \!\in \!\mathcal{U}$), and let $A_{i}(t)$ denote the number of arrival packets during slot $t$.
	For the PTQ in both SemCom and BitCom cases, its packet departure rate depends on the number of packets sent out from MU $i$ to BS $j$ ($\forall j \!\in \!\mathcal{J}$) during slot $t$, denoted by $D_{ij}\left(t\right)$, which has the probability mass function (PMF) as
	\begin{equation}
		\label{PTQDepart}
		\begin{aligned}
			\Pr&\left\{D_{ij}\left(t\right)=k\right\}=\Pr\left\{\left\lfloor \frac{Tz_{ij}\log_{2}\left( 1+\gamma_{ij}\left(t\right)\right)}{L}\right\rfloor =k\right\}\\
		\end{aligned}
	\end{equation}
	Here, $\left\lfloor\cdot\right\rfloor$ is the floor function, and $L$ denotes size of each packet in bits, $k=0, 1, 2, \cdots$.
	Given any reasonable probability distribution of SINR $\gamma_{ij}\left(t\right)$, applying its cumulative distribution function (CDF) directly yields the close-form expression of \eqref{PTQDepart} can be obtained directly~\cite{moustakas2013sinr}.
	Besides, assuming the distribution of $\gamma_{ij}\left(t\right)$ keeps the same with any time slot $t$.

	Notably, each packet generated at a SemCom-enabled MU requires a certain type of background knowledge~\cite{10261329}, hence resulting in either a knowledge-matching or -mismatching state with its receiver.
	For illustration, let $I_{i}^{\mathit{Mat}}$ denote the semantic-coding time required by a knowledge-matching packet with mean $1/\mu_{i}^{\mathit{Mat}}$ (in \textit{s/packet}), and let $I_{i}^{\mathit{Mis}}$ denote the semantic-coding time required by a knowledge-mismatching packet with mean $1/\mu_{i}^{\mathit{Mis}}$ ($\mu_{i}^{\mathit{Mat}}\!>\!\mu_{i}^{\mathit{Mis}}$ in practice).
	Without loss of generality, $I_{i}^{\mathit{Mat}}$ and $I_{i}^{\mathit{Mis}}$ are assumed to be two exponential random variables independent of each other.
	
	Having these, the overall service time distribution of packets at SCQ should be treated as a general distribution~\cite{xia2023xurllc}, and thus the SCQ fully qualiﬁes as an M/G/1 model.
	Since the average proportion of knowledge-matching packets to the total packets in the SCQ can be inscribed by the average knowledge-matching degree $\tau_{i}$ of MU $i$, the average semantic-coding time of a packet at SCQ becomes $I_{i}=\tau_{i}I_{i}^{\mathit{Mat}}+\left(1-\tau_{i}\right)I_{i}^{\mathit{Mis}}$.
	As such, the average packet queuing latency of SCQ, denoted by $\delta_{i}^{\mathit{S_1}}$, can be calculated via applying the~\textit{Pollaczek-Khintchine formula}~\cite{ross2014introduction},\footnote{A prerequisite of $\lambda_{i}\mathds{E}\left[I_{i}\right]\!< \!1$ is assumed to guarantee the steady state for the M/G/1 system and make its queuing latency finite and solvable~\cite{ross2014introduction}.} as shown in~\eqref{SCQ_queuing_delay} at the bottom of this page.
	\begin{figure*}[hb]
		\centering
		\hrulefill
		\begin{equation}
		\label{SCQ_queuing_delay}
			\begin{aligned}
				\delta_{i}^{\mathit{S_1}}&\!=\!\frac{\lambda_{i}\left(\mathds{E}^{2}\left[I_{i}\right]+\mathds{V}\left(I_{i}\right)\right)}{2\left(1-\lambda_{i}\mathds{E}\left[I_{i}\right]\right)}\!+\!\mathds{E}\left[I_{i}\right]\!=\! \frac{\lambda_{i}\left[\tau_{i}\left(1-\tau_{i}\right)/\mu_{i}^{\mathit{Mat}}\mu_{i}^{\mathit{Mis}}\!+\!\left(\tau_{i}/\mu_{i}^{\mathit{Mat}}\right)^2\!+\!\left(\left(1-\tau_{i}\right)/\mu_{i}^{\mathit{Mis}}\right)^2\right]}{1-\lambda_{i}\tau_{i}/\mu_{i}^{\mathit{Mat}}-\lambda_{i}\left(1-\tau_{i}\right)/\mu_{i}^{\mathit{Mis}}}+\frac{\tau_{i}}{\mu_{i}^{\mathit{Mat}}}+\frac{1-\tau_{i}}{\mu_{i}^{\mathit{Mis}}}.
			\end{aligned}
		\end{equation}
	\end{figure*}
	
	As for the number of packets arriving at the SemCom-enabled PTQ in slot $t$, denoted by $A_{i}'(t)$, it should exactly be the number of packets leaving its tandem SCQ in the same slot, according to the defined two-stage tandem structure.
	Meanwhile, it is not difficult to deduce that the knowledge-matching packets leaving the SCQ follow a Poisson process with mean $\mu_{i}^{\mathit{Mat}}$, while the knowledge-mismatching packets leave as a Poisson process with mean $\mu_{i}^{\mathit{Mis}}$.
	As such, $A_{i}'(t)$ should still satisfy the Poisson distribution with mean $\lambda_{i}'=\tau_{i}\mu_{i}^{\mathit{Mat}}+(1-\tau_{i})\mu_{i}^{\mathit{Mis}}$.
	In addition, we assume that in any $t$, the packets to be transmitted leave the queue first and then the arriving packets enter it.
	If let $Q_{ij}\left(t\right)$ denote its queue length for the link between MU $i$ and BS $j$ at slot $t$, $t=1,2,\cdots,N-1$, we have $Q_{ij}\left(t+1\right)\triangleq \min\left\{\max\left\{Q_{ij}\left(t\right)\!-\!D_{ij}\left(t\right),0\right\}\!+\!A'_{i}\!\left(t\right),F\right\}$.
		Given any queue length $a$ at slot $t$ (i.e., $Q_{ij}\left(t\right)=a$, $0\leqslant a\leqslant F$), $Q_{ij}\left(t+1\right)$ is determined only by $A'_{i}\left(t\right)$ and $D_{ij}\left(t\right)$.
		Apparently, the stochastic $Q_{ij}\left(t\right)$ across all slots forms a discrete-time Markov process, herein we define $\omega^{a\looparrowright b}_{ij}\left(t\right)=\Pr\left\{Q_{ij}\left(t+1\right)=b\mid Q_{ij}\left(t\right)=a\right\}$ as its one-step state transition probability from length $a$ to length $b$ at slot $t+1$, $0\leqslant b\leqslant F$.
		Since the PMFs of both $A'_{i}\left(t\right)$ and $D_{ij}\left(t\right)$ are independent of $t$, we re-denote them by $A'_{i}$ and $D_{ij}$ for brevity, respectively.
		As such, $\omega^{a\looparrowright b}_{ij}\left(t\right)$ can be expressed as $\omega^{a\looparrowright b}_{ij}$ as well.
		Having these, we have the one-step state transition probability matrix of SemCom-enabled PTQ, denoted as $\bm{\Omega}_{ij}=\left(\omega^{a\looparrowright b}_{ij}\right)_{0\leqslant a,b \leqslant F}$,
		where each $\omega^{a\looparrowright b}_{ij}$ can be calculated by simultaneously taking into account the PMFs of $A'_{i}$ and $D_{ij}$ over different values of $a$ and $b$.
		According to~\cite{guo2019resource}, there exists a unique steady-state probability vector $\bm{\alpha}_{ij}=\left[\alpha_{ij}^{0},\alpha_{ij}^{1},\cdots,\alpha_{ij}^{F}\right]^{T}$, which can be obtained by solving $\bm{\Omega}_{ij}^{T}\bm{\alpha}_{ij}=\bm{\alpha}_{ij}$ and $\sum_{k=0}^{F}\alpha_{ij}^{k}=1$.
	Therefore, the long-term average queue length of $Q_{ij}(t)$ can be obtained by $\mathds{E}\left[Q_{ij}(t)\right]=\sum_{k=0}^{F}k\alpha_{ij}^{k}$.
	Moreover, by combining $\bm{\alpha}_{ij}$ with the PMFs of PTQ's packet arrival and departure, the average number of packets dropped at the steady-state PTQ during any slot $t$, denoted by $G_{ij}$, can be calculated by \eqref{Drop}, as shown at the bottom of the next page.
	As its average total packet arrival rate is $\lambda_{i}'$, we can obtain the steady-state average packet loss ratio of SemCom-enabled PTQ to represent the proportion of packets failed to be delivered to all arriving packets as
	\begin{figure*}[hb]
		\centering
		\hrulefill
		\begin{equation}
				\begin{aligned}
				\label{Drop}
					G_{ij}\!=\!\sum_{f=1}^{F}\Pr\!\left\{\!A_{i}'\!=\!f\!\right\}\!\left[\sum_{k=0}^{f-1}\Pr\left\{D_{ij}\!\leqslant\! k\right\}\!\left(1\!-\!W_{ij}^{(F-f+k)}\right)\!\right] +\! \!\sum_{f\!=\!F+1}^{\infty}\!\!\!\Pr\!\left\{\!A_{i}'\!=\!f\!\right\}\!\left[\!(f\!-\!F)\!+\! \!\sum_{k=0}^{F-1}\!\Pr\!\left\{D_{ij}\!\leqslant\! k\right\}\!\left(\!1\!-\!W_{ij}^{(k)}\!\right)\!\right].
				\end{aligned}
			\end{equation}
	\end{figure*}
	\begin{equation}
	\label{SemComplr}
		\theta_{ij}^{\mathit{S}}=\frac{G_{ij}}{\lambda_{i}'T}=\frac{G_{ij}}{\tau_{i}\mu_{i}^{\mathit{Mat}}T+\mu_{i}^{\mathit{Mis}}T-\tau_{i}\mu_{i}^{\mathit{Mis}}T}.
	\end{equation}
	As such, the average effective packet arrival rate becomes $\lambda_{i}^{\mathit{eff}}=\left(1-\theta_{ij}^{\mathit{S}}\right)\lambda_{i}'=\tau_{i}\mu_{i}^{\mathit{Mat}}+(1-\tau_{i})\mu_{i}^{\mathit{Mis}}-G_{ij}/T$.
	In this way, we can apply Little's law~\cite{little2008little} to finalize the steady-state average queuing latency of SemCom-enabled PTQ as
	\begin{equation}
		\label{PTQ_queuing_delay}
		\begin{aligned}
		\delta_{ij}^{\mathit{S_2}}=\frac{\mathds{E}\left[Q_{ij}(t)\right]}{\lambda_{i}^{\mathit{eff}}}=\frac{\sum_{k=0}^{F}k\alpha_{ij}^{k}}{\tau_{i}\mu_{i}^{\mathit{Mat}}+(1-\tau_{i})\mu_{i}^{\mathit{Mis}}-G_{ij}/T}.
		\end{aligned}
	\end{equation}
	Combined with SemCom-enabled SCQ's queuing latency $\delta_{i}^{\mathit{S_1}}$, we obtain the overall average queuing latency of the link between SemCom-enabled MU $i$ and BS $j$ as $\delta_{ij}^{S}=\delta_{i}^{\mathit{S_1}}+\delta_{ij}^{\mathit{S_2}}$.
	
	Similarly, the average packet loss ratio and queuing latency of BitCom-enabled PTQ are then denoted as $\theta_{ij}^{\mathit{B}}$ and $\delta_{ij}^{\mathit{B}}$, respectively.
	The same mathematical methods as the above can be employed, where only the PMF and mean of $A'_{i}(t)$ in each relevant term need to be substituted with that of $A_{i}(t)$.
	If considering both the SemCom case and the BitCom case, the average queuing latency experienced by the link should be $\delta_{ij}=y_{ij}\delta_{ij}^{S}+(1-y_{ij})\delta_{ij}^{B}$, and the average packet loss ratio that indicates the communication reliability of the link should be found by $\theta_{ij}=y_{ij}\theta_{ij}^{S}+(1-y_{ij})\theta_{ij}^{B}$.

\subsection{Problem Formulation}
	Let us define three variable sets $\bm{x}=\left\{x_{ij}\mid i \in \mathcal{U},j \in \mathcal{J}\right\}$, $\bm{y}=\left\{y_{ij}\mid i \in \mathcal{U},j \in \mathcal{J}\right\}$, and $\bm{z}=\left\{z_{ij}\mid i \in \mathcal{U}, j \in \mathcal{J}\right\}$ that consist of all possible indicators pertinent to UA, MS, and BA, respectively.
	With the aim of maximizing the overall message throughput of the HSB-Net, our joint optimization problem is
	\begin{align}
	\mathbf{P1}:\ \max_{\bm{x},\bm{y},\bm{z}} \quad & \sum_{i\in \mathcal{U}}\sum_{j\in \mathcal{J}}x_{ij}\overline{M}_{ij}~\label{P1}\\
	{\rm s.t.} \quad & \sum_{j\in \mathcal{J}} x_{ij}= 1,\ \forall i\in \mathcal{U},\tag{\ref{P1}a}\\
	& \sum_{i\in \mathcal{U}}x_{ij} z_{ij}\leqslant Z_{j},\ \forall j\in \mathcal{J},\tag{\ref{P1}b}\\
	& x_{ij}\delta_{ij}\leqslant \delta_{0},\ \forall \left( i,j\right) \in\mathcal{U}\times \mathcal{J},\tag{\ref{P1}c}\\
	& x_{ij}\theta_{ij}\leqslant \theta_{0},\ \forall \left( i,j\right) \in\mathcal{U}\times \mathcal{J},\tag{\ref{P1}d}\\
	& \sum_{j\in \mathcal{J}}x_{ij}\overline{M}_{ij}\geqslant \mathit{M}_{i}^{o},\ \forall i\in \mathcal{U},\tag{\ref{P1}e}\\
	& x_{ij},y_{ij}\!\in \!\left\{ 0,1\right\},\ \forall \left( i,j\right) \in\mathcal{U}\times \mathcal{J}.\tag{\ref{P1}f}
	\end{align}
	Constraints (\ref{P1}a) and (\ref{P1}b) model the single-BS constraint for UA and the bandwidth budget for BA, respectively.
	Constraints (\ref{P1}c) and (\ref{P1}d) define the average packet queuing latency and loss ratio requirements for each associated link.
	$M_{i}^{th}$ in constraint (\ref{P1}e) represents a minimum message throughput threshold for each MU $i$'s association link, while constraint (\ref{P1}f) characterizes the binary properties of both $\bm{x}$ and $\bm{y}$.
	As such, the main difficulty in solving $\mathbf{P1}$ lies in the combinatorial nature of variables $\bm{x}$ and $\bm{y}$ coupled with the non-convexity of its highly complicated constraints (\ref{P1}c) and (\ref{P1}d).

	\section{Optimal Resource Management in HSB-Nets}
	
	\subsection{Strategy Determination for UA and MS}
	Assume that there is a minimum bandwidth amount that BS $j$ should allocate to each MU $i$, in order to simultaneously meet the preset latency, reliability, and message throughout requirements.
	Intuitively, the steady-state average packet queuing latency $\delta_{ij}$ and the steady-state average packet loss ratio $\theta_{ij}$ should be monotonically non-increasing w.r.t. $z_{ij}$ with any $y_{ij}$.
	Proceeding as in~\cite{10261329}, $\Re_{ij}(\cdot)$ is known to be a monotonically increasing function of $z_{ij}$, and thus $\overline{M}_{ij}$ should also monotonically increase w.r.t. $z_{ij}$ in either the case of $y_{ij}=0$ or $y_{ij}=1$.
	As such, by separately considering the boundary situations of constraints (\ref{P1}c)-(\ref{P1}e), we can obtain three minimum bandwidth thresholds for the SemCom case (denoted as $z_{ij}^{S_{\delta}}$, $z_{ij}^{S_{\theta}}$, and $z_{ij}^{S\!_{M}}$) and three thresholds for the BitCom case (denoted as $z_{ij}^{B_{\delta}}$, $z_{ij}^{B_{\theta}}$, and $z_{ij}^{B\!_{M}}$), respectively.
	
	
	Now, we aim at the optimal $\bm{x}^{*}=\left\{x_{ij}^{*}\mid i \in \mathcal{U}, j\in \mathcal{J}\right\}$ and $\bm{y}^{*}=\left\{y_{ij}^{*}\mid i \in \mathcal{U}, j\in \mathcal{J}\right\}$ by fixing each SemCom-associated $z_{ij}$ term as $z_{ij}^{S_{th}}\!=\!\max\!\left\{\!z_{ij}^{S\!_{M}}\!,z_{ij}^{S_{\delta}}\!,z_{ij}^{S_{\theta}}\!\right\}$ and each BitCom-associated $z_{ij}$ as $z_{ij}^{B_{th}}\!=\!\max\!\left\{\!z_{ij}^{B\!_{M}}\!,z_{ij}^{B_{\delta}}\!,z_{ij}^{B_{\theta}}\!\right\}$.
	Then, constraints (\ref{P1}c)-(\ref{P1}e) can be all removed, and $\mathbf{P1}$ becomes
	\begin{align}
		\mathbf{P1.1}:\ \max_{\bm{x},\bm{y}} \quad & \sum_{i\in \mathcal{U}}\sum_{j\in \mathcal{J}}x_{ij}\!\left[y_{ij}\overline{M}_{ij}^{S_{th}}\!+\!\left(1\!-\!y_{ij}\right)\!\overline{M}_{ij}^{B_{th}}\!\right]~\label{P1.1}\\
		{\rm s.t.} \quad & \sum_{i\in \mathcal{U}}\!x_{ij}\!\left[y_{ij}z_{ij}^{S_{th}}\!+\!\left(1\!-\!y_{ij}\right)z_{ij}^{B_{th}}\!\right]\!\leqslant \!Z_{j},\tag{\ref{P1.1}a}\\
		& \text{(\ref{P1}a)},\ \text{(\ref{P1}f)},\tag{\ref{P1.1}b}
	\end{align}
	where let $\overline{M}_{ij}^{S_{th}}\!=\! \tau_{i}\Re_{ij}\!\left(\!z_{ij}^{S_{th}}\!\log_{2}\!\left(1+\overline{\gamma}_{ij}\right)\!\right)$ and $\overline{M}_{ij}^{B_{th}}\!\!=\!\!\rho_{ij}z_{ij}^{B_{th}}\!\log_{2}\!\left( 1+\overline{\gamma}_{ij}\right)$, both are regarded as known constants.
	
	Regarding $\mathbf{P1.1}$, we incorporate constraint (\ref{P1.1}a) into its objective function (\ref{P1.1}) by associating Lagrange multipliers $\bm{\eta}=\left\{\eta_{j}\mid \eta_{j}\geqslant 0, j \in \mathcal{J}\right\}$.
	In this way, the Lagrange dual problem of $\mathbf{P1.1}$ becomes
	\begin{align}
			\mathbf{D1.1}:\ \min_{\bm{\eta}} \quad & F\left(\bm{\eta}\right)=g_{\bm{x},\bm{y}}\left(\bm{\eta}\right)+\sum_{j\in \mathcal{J}}\eta_{j}Z_{j}~\label{D},
	\end{align}
	where
	\begin{equation}
		\label{Dual}
			\begin{split}
			g_{\bm{x},\bm{y}}\left(\bm{\eta}\right) \ &= \ \sup_{\bm{x},\bm{y}} \ \sum_{i\in \mathcal{U}}\sum_{j\in \mathcal{J}}\bigl[x_{ij}y_{ij}\!\left(\overline{M}_{ij}^{S_{th}}\!-\!\eta_{j}z_{ij}^{S_{th}}\!\right)\\
			& \qquad \qquad +x_{ij}\left(1\!-\!y_{ij}\right)\!\left(\overline{M}_{ij}^{B_{th}}-\eta_{j}z_{ij}^{B_{th}}\!\right)\bigl]\\
			{\rm s.t.} \ & \ \text{(\ref{P1}a)},\ \text{(\ref{P1}f)}.
		    \end{split}
	\end{equation}
	
	Next, we extend the original BS indicator set $\mathcal{J}$ to $\mathcal{J}'=\{1,2,\cdots,J,J+1,J+2,\cdots,2J\}$ and define a new set of variables $\bm{\nu}=\{\nu_{ij'}\mid i \in \mathcal{U}, j\in \mathcal{J}'\}$ w.r.t. \eqref{Dual} as
	\begin{equation}
		\label{newvariableset}
			\nu_{ij'}=\begin{cases}
			x_{ij'}y_{ij'}, & \text{if } j' \in \mathcal{J};\\
			x_{i(j'-J)}(1-y_{i(j'-J)}),& \text{if } j' \in \mathcal{J}'\backslash \mathcal{J}.
		\end{cases}
	\end{equation}
	In parallel, we define another new set of constants $\bm{\xi}=\{\xi_{ij'}\mid i \in \mathcal{U}, j\in \mathcal{J}'\}$ to characterize the coefficient of each $\nu_{ij'}$, i.e.,
	\begin{equation}
		\label{newconstantset}
			\xi_{ij'}=\begin{cases}
			\overline{M}_{ij'}^{S_{th}}\!-\!\eta_{j'}z_{ij'}^{S_{th}},& \text{if } j' \in \mathcal{J};\\
			\overline{M}_{i(j'-J)}^{B_{th}}\!-\!\eta_{(j'-J)}z_{i(j'-J)}^{B_{th}},& \text{if } j' \in \mathcal{J}'\backslash \mathcal{J}.
		\end{cases}
	\end{equation}
	As such, given the initial dual variable $\bm{\eta}$, problem \eqref{Dual} should be converted to
	\begin{align}
		\mathbf{P1.2}:\ \max_{\bm{\nu}} \quad & \sum_{i\in \mathcal{U}}\sum_{j'\in \mathcal{J'}}\xi_{ij'}\nu_{ij'}~\label{P1.2}\\
		{\rm s.t.} \quad & \sum_{j'\in \mathcal{J}'}\nu_{ij'}= 1,\ \forall i\in \mathcal{U},\tag{\ref{P1.2}a}\\
		& \nu_{ij'}\in \left\{ 0,1\right\},\ \forall \left( i,j'\right) \in\mathcal{U}\times \mathcal{J}'.\tag{\ref{P1.2}b}
	\end{align}
	It is easily derived from $\mathbf{P1.2}$ that for any $i\in \mathcal{U}$, the optimal $j'$ satisfies $\widehat{j'}=\arg \max_{j'\in \mathcal{J}'}\xi_{ij'}, \forall i \in \mathcal{U}$.
	Then, we determine $\bm{x}^{*}$ and $\bm{y}^{*}$ for each MU $i$ and BS $j$ in the HSB-Net by
	\begin{equation}
		\label{eachiteration}
		\begin{cases}
			x_{ij}^{*}=1,\ y_{ij}^{*}=1,& \text{if } \widehat{j'} \in \mathcal{J} \text{ and } j=\widehat{j'};\\
			x_{ij}^{*}=1,\ y_{ij}^{*}=0,& \text{if } \widehat{j'}\!\in\! \mathcal{J}'\backslash \mathcal{J} \text{ and } j\!=\!\widehat{j'}\!-\!J;\\
			x_{ij}^{*}\!=\!0,& \text{otherwise}.
		\end{cases}
	\end{equation}
	
	Afterward, the partial derivatives w.r.t. $\bm{\eta}$ in the objective $F\left(\bm{\eta}\right)$ in $\mathbf{D1.1}$ are set as the subgradient direction in each update iteration.
	Now suppose that in a certain iteration, say iteration $l$, each $\eta_{j}$ ($j \in \mathcal{J}$) is updated as the following rule:
	\begin{equation}
		\label{multiplierupdate}
		\eta_{j}\left(l+1\right)=\max\left\{\eta_{j}\left(l\right)-\epsilon(l)\cdot\nabla F\left(\eta_{j}\right),0\right\},
	\end{equation}
	where $\nabla F\left(\eta_{j}\right)=Z_{j}-\sum_{i\in \mathcal{U}}x_{ij}\left[y_{ij}z_{ij}^{S_{th}}+\left(1-y_{ij}\right)z_{ij}^{B_{th}}\right]$
	and $\epsilon\left(l\right)$ denotes the stepsize of update in iteration $l$.
	In general, the convergence of the subgradient descent method can be guaranteed with a properly preset stepsize~\cite{jiang2016optimal}.
	
	Nevertheless, it is worth noting that the above solutions cannot always directly reach the optimality of $\mathbf{P1.1}$, as the BA constraint (\ref{P1.1}a) may be violated at some BSs within each iteration.
	In this case, here we additionally adopt a preference list-based heuristic algorithm to project the solution obtained in each iteration back to the feasible set of (\ref{P1.1}a).
	That is, for each BS that violates (\ref{P1.1}a), we choose to reallocate its associated MUs who are consuming the most bandwidth to other BSs according to (\ref{eachiteration}), until the bandwidth constraints are fulfilled at all BSs after each iteration.

	\subsection{Optimal Solution for BA}
	According to the obtained $\bm{x}^{*}$ and $\bm{y}^{*}$, we aim to reallocate all bandwidth resources of each BS $j$ ($\forall j \in \mathcal{J}$) to all its associated MUs, thus a total of $S$ BA subproblems w.r.t. $\mathbf{P1}$ are constructed.
	Given the preset $z_{ij}^{S_{th}}$ and $z_{ij}^{B_{th}}$, each BA subproblem of BS $j$ is formulated as follows:
	\begin{align}
		\mathbf{P1.3}_{j}:\ \max_{\bm{z}} \quad & \sum_{i\in \mathcal{U}_{j}^{S}}\overline{M}_{ij}^{S}+\sum_{i\in \mathcal{U}_{j}^{B}}\overline{M}_{ij}^{B}~\label{P1.3}\\
		{\rm s.t.} \quad & \sum_{i\in \mathcal{U}_{j}^{S}\cup\mathcal{U}_{j}^{B}}z_{ij}=Z_{j},\tag{\ref{P1.3}a}\\
		& z_{ij}\geqslant z_{ij}^{S_{th}},\ \forall i \in\mathcal{U}_{j}^{S},\tag{\ref{P1.3}b}\\
		& z_{ij}\geqslant z_{ij}^{B_{th}},\ \forall i \in\mathcal{U}_{j}^{B},\tag{\ref{P1.3}c}
	\end{align}
	where $\mathcal{U}_{j}^{S}\!=\!\left\{i\mid i\in\mathcal{U},x_{ij}^{*}=1,y_{ij}^{*}=1\right\}$ stands for the set of all SemCom-enabled MUs associated with BS $j$, and $\mathcal{U}_{j}^{B}=\left\{i\mid i\in\mathcal{U},x_{ij}^{*}=1,y_{ij}^{*}=0\right\}$ indicates the set of all BitCom-enabled MUs associated with BS $j$.
	Then given the convex property of $\Re_{ij}(\cdot)$~\cite{10261329}, we have the objective function and all constraints of each $\mathbf{P1.3}_{j}$ are convex, thereby some efficient optimization toolboxes such as CVPXY~\cite{diamond2016cvxpy} can be directly applied to finalize the optimal BA solution for the HSB-Net.
	
	
	\section{Numerical Results and Discussions}
	We consider a total of $200$ MUs and $10$ BSs randomly dropped in a circular area with a radius of $300$ meters, where each SINR $\gamma_{ij}$ follows a Gaussian distribution with standard deviation of $4$ dB~\cite{moustakas2013sinr}.
	Specifically, each MU has a transmit power of $20$ dBm and each BS has a bandwidth budget of $15$ MHz~\cite{xu2017power}.
	The time slot length $T$ is set to $1$ ms, the size of each packet $L$ is $800$ bits, and the buffer size of PTQ $F$ is $20$.
	
	In SemCom-relevant settings, we put all MUs into a wireless text communication scenario, in which the implementation of SemCom has been well studied and the Transformer is adopted as the uniﬁed semantic encoder to realize $\Re_{ij}(\cdot)$ for each SemCom link~\cite{10261329}.
	As for the queuing modeling part, the average knowledge-matching degree $\tau_{i}$, message throughput threshold $\mathit{M}_{i}^{o}$, and BitCom-based B2M coefficient $\rho_{ij}$ are randomly generated for each MU in the range of $0.6 \sim 1$, $50 \sim 100$, and $2\times 10^{-5}\sim 2\times 10^{-4}$, respectively.
	Besides, the average packet arrival rate $\lambda_{i}$ is prescribed at $1000$ packets/s, while the average interpretation times of knowledge-matching and -mismatching packets in SCQ are considered as $8\times 10^{-4}$ and $1\times 10^{-3}$ s/packet, respectively.
	Furthermore, we set $\delta_{0}$ as $20$ ms, $\theta_{0}$ as $0.01$, and the dynamic stepsize of $\epsilon(l)= 1\times 10^{-6}/l$ to update the Lagrange multipliers in~(\ref{multiplierupdate}).
	
	For comparison purposes, the max-SINR UA scheme is employed combined with two heuristic MS schemes and two typical BA schemes as four different benchmarks: (MS-I) A~\textit{knowledge matching degree-based} method, where MU $i$ selects the SemCom mode if $\tau_i$ is above a preset threshold of $0.8$, and otherwise selects the BitCom mode; (MS-II) A~\textit{SINR-based} method, where MU $i$ selects the BitCom mode if $\gamma_{ij}$ is above a preset threshold of $6$ dB, and otherwise selects the SemCom mode; (BA-I) The~\textit{water-filling} algorithm~\cite{he2013water}; (BA-II) The~\textit{evenly-distributed} algorithm~\cite{ye2013user}.

	\begin{figure}[t]
		\centering
		\includegraphics[width=0.4\textwidth]{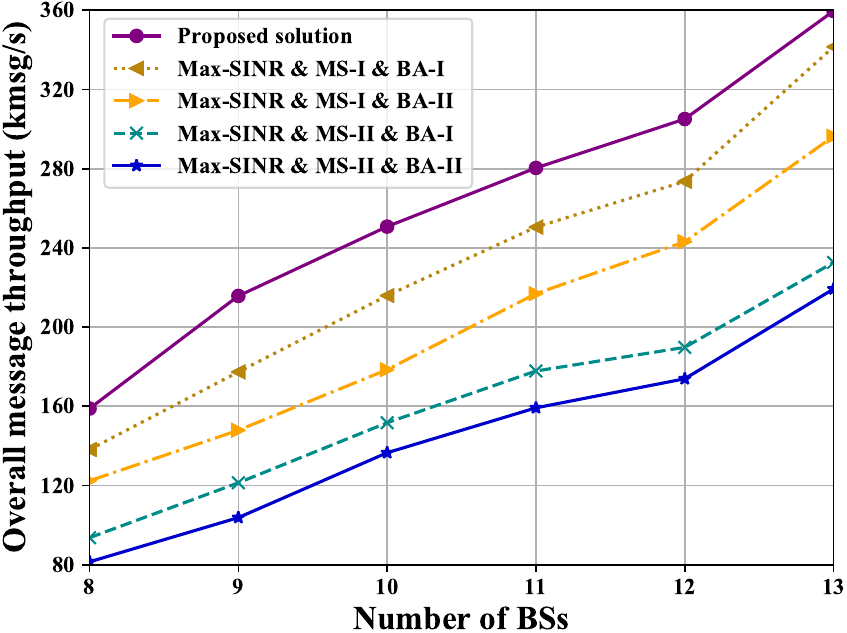} 
		\caption{Overall message throughput (\textit{kmsg/s}) versus different numbers of BSs.}
		\label{SimuFigure4}
    \end{figure}
    
\begin{figure}[t]
		\centering
		\includegraphics[width=0.4\textwidth]{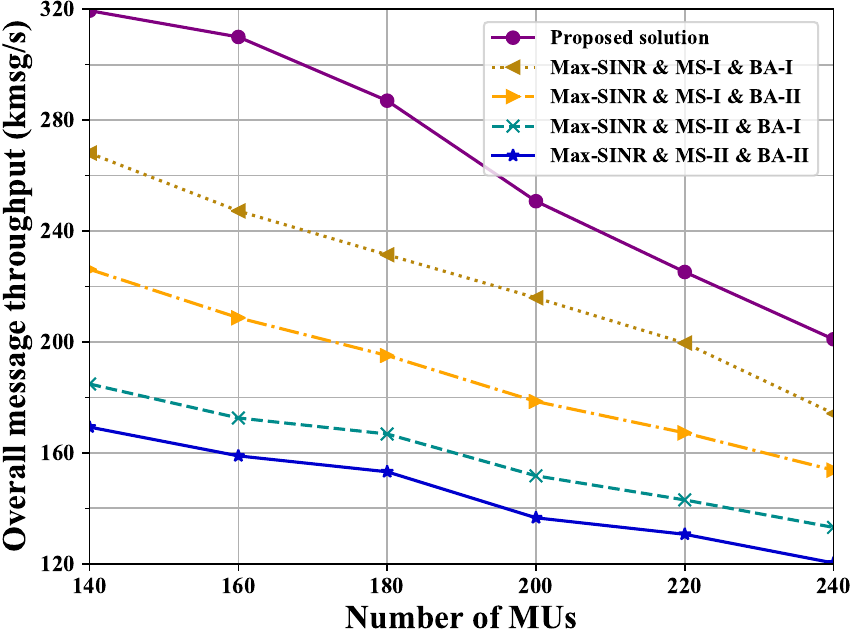} 
		\caption{Overall message throughput versus different numbers of MUs.}
		\label{SimuFigure5}
    \end{figure}

    To validate our proposed solution, we first test the overall message throughput under different numbers of BSs and MUs in Fig.~\ref{SimuFigure4} and Fig.~\ref{SimuFigure5}, respectively, compared with the four benchmarks.
    As elucidated in Fig.~\ref{SimuFigure4}, by varying the number of BSs from $8$ to $13$, the message throughput performance of the proposed solution gradually increases from around $160$ to $360$ kmsg/s, and consistently outperforms these benchmarks.
    For example, a performance gain of the proposed solution is about $29.9$ kmsg/s compared with the benchmark of Max-SINR plus (MS-I) plus (BA-I) and $102.6$ kmsg/s compared with the benchmark of Max-SINR plus (MS-II) plus (BA-I) when $11$ BSs are set up in the HSB-Net.
    Here, such uptrend is apparent since more BSs represent that more bandwidth resources are available for MUs to achieve higher message rates.
    Particularly in such an uplink scenario of HSB-Net, the increase in the number of BSs does not have any influence on channel interference, hence a stable growth can be observed.
    
    Fig.~\ref{SimuFigure5} demonstrates a downward trend of message throughput performance when rising the number of MUs from $140$ to $240$.
    To be concrete, the overall network performance is already saturated at the very beginning in holding $140$ MUs and then deteriorates with the addition of MUs, as the effect of severe channel interference from excessive MUs starts to dominate the more availability of resources.
	In the meantime, it can be seen that our solution still surpasses all the four benchmarks with a significant performance gain.
    For instance, with $160$ MUs in the HSB-Net, the proposed solution realizes a message throughput of about $310$ kmsg/s, i.e., $1.5$ times that of the Max-SINR plus (MS-I) plus (BA-II) scheme and $2$ times that of the Max-SINR plus (MS-II) plus (BA-II) scheme.

    \begin{figure}[t]
		\centering
		\includegraphics[width=0.4\textwidth]{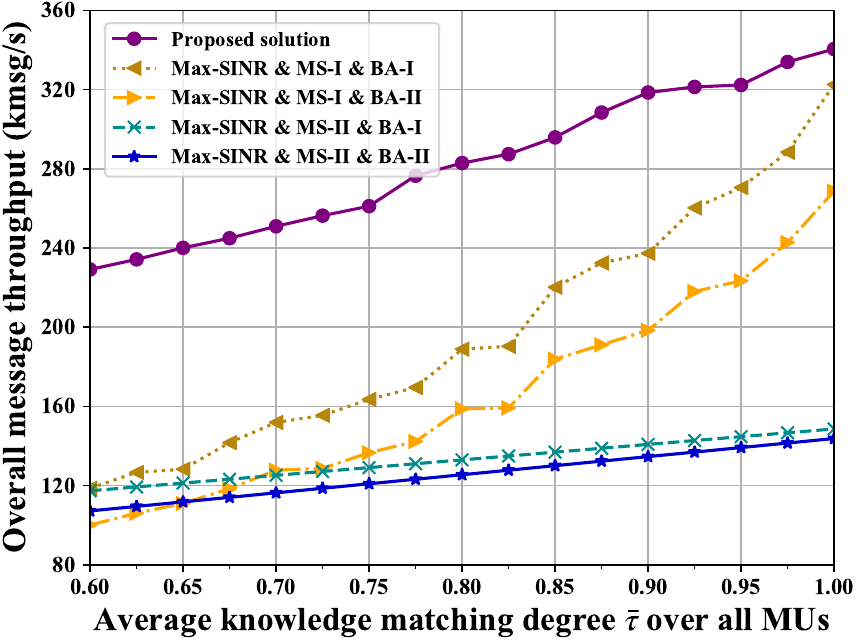} 
		\caption{Overall message throughput versus varying $\bar{\tau}$ over all $200$ MUs.}
		\label{SimuFigure6}
    \end{figure}
    
    In addition, we compare the message throughput performance with varying overall average knowledge-matching degree $\bar{\tau}=\frac{1}{U}\sum_{i \in \mathcal{U}}\tau_{i}$ as shown in Fig.~\ref{SimuFigure6}.
    Again, our solution still outperforms these benchmarks with the considerable performance gain, especially in the low $\bar{\tau}$ region.
    Besides, a growing message throughput is observed by all solutions as $\bar{\tau}$ increases, and our solution and the (MS-I) scheme are more affected by changes in $\bar{\tau}$ compared to the (MS-II).
    This is first due to the message-throughput-priority design in our objective function~(\ref{P1}), therefore, our solution is more likely to generate more SemCom-enabled MUs with larger $\bar{\tau}$.
    Likewise, more SemCom-enabled MUs can exist in the same case according to the prescribed (MS-I) scheme, while the number of SemCom-enabled MUs is only affected by SINR in (MS-II), and thus roughly keeps stable irrespective of the change in $\bar{\tau}$.
    
     \begin{figure}[t]
		\centering
		\includegraphics[width=0.4\textwidth]{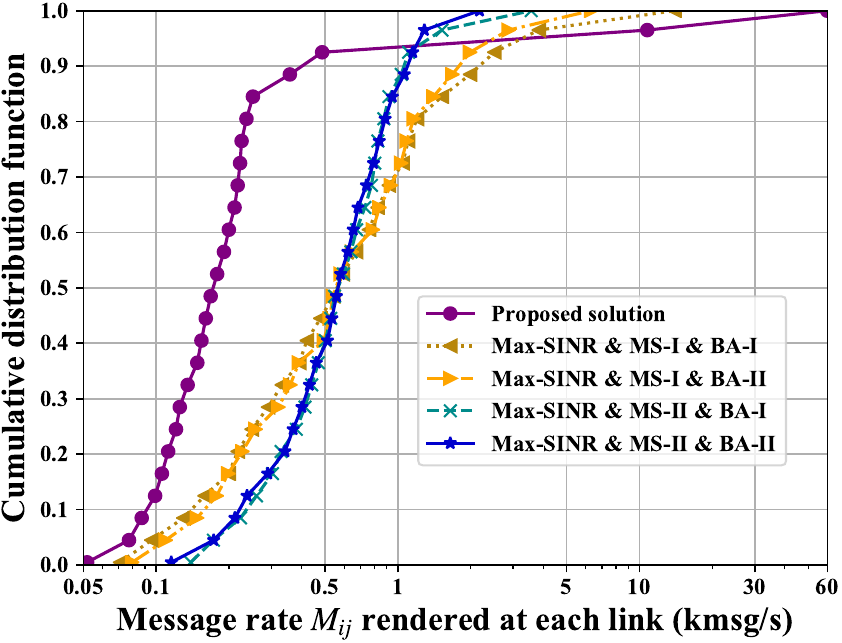} 
		\caption{The CDF of the message rate $M_{ij}$ obtained by the associated link.}
		\label{SimuFigure7}
    \end{figure}
	
	Finally, the CDFs of the message rate $M_{ij}$ rendered at all links are plotted in Fig.~\ref{SimuFigure7}.
	Although most MUs in our solution only get the lower message rates compared with these benchmarks, this is quite reasonable since our optimization to $\mathbf{P1}$ focuses on the maximization for overall message throughput of all MUs in the HSB-Net.
	Hence, it can be interpreted as that the proposed solution choose to sacrifice user semantic fairness in favor of devoting more bandwidth resources to a smaller number of MUs with better average knowledge-matching degrees, B2M transformation, and SINRs.
	
	\section{Conclusions}
	In this paper, we investigated the resource management problem in the novel and practical HSB-Net scenario, where SemCom and BitCom modes coexist.
	First, a B2M transformation function was introduced to identify the message throughput of each link.
	We then modelled a two-stage tandem queue model for semantic packet processing and transmission, and derived the average packet loss ratio and queuing latency.
	On this basis, a joint optimization problem was formulated to maximize the overall message throughput of HSB-Net, and we utilized a Lagrange primal-dual method alongside a preference list-based heuristic algorithm to seek the optimal UA, MS, and BA solutions.
	Numerical results validated the performance superiority of the proposed solution in terms of overall message throughput compared with benchmarks.

	\bibliographystyle{IEEEtran}
	\bibliography{main}

\end{document}